# BPZT HBARs for bias-tunable, stress generation at GHz frequencies

U.K. Bhaskar,[1]* D. Tierno,[2] G. Talmelli,[2] F. Ciubotaru,[2] C. Adelmann,[2] and T. Devolder[1]
[1] Centre de Nanosciences et de Nanotechnologies, CNRS, Université Paris-Sud, Université Paris-Saclay, 91120 Palaiseau, France
[2]Imec, 3001 Leuven, Belgium

*Abstract*— **The high frequency performance of strong piezoelectric materials like PZT remains relatively less explored due to the assumption of large dielectric/ferroelectric losses at GHz frequencies. Recently, the advent of magnetoelectric technology as an on-chip route to excite magnetization dynamics has provided the impetus to evaluate the electromechanical performance of PZT at microwave frequencies. In this work, we demonstrate that HBARs fabricated using Barium-doped PZT (BPZT) films can efficiently generate acoustic waves up to 15 GHz. The ferroelectricity of BPZT endows added functionality to the resonator in the form of voltage tunability of the electromechanical performance. We extract the piezoelectric coefficient by numerically comparing the performance of BPZT with the Mason model. The extracted piezoelectric coefficient ~60 pm/V agrees well with reported values on thin film PZT measured at low frequencies (<100 MHz). Our results suggest that with further improvement in device design and material processing, BPZT resonators could operate as large amplitude, tunable stress transducers at GHz frequencies.**

## I. Introduction

Strain engineering has empowered the semiconductor revolution by providing tools to enhance the mobility of transistors[1], and control the bandgap of heterostructures[2]. The role of strain in microelectronics is set to become more significant with recent proposals involving the incorporation of magnetic logic devices[3],[4]. Indeed, mechanical deformation is widely considered as a technologically relevant, energy-efficient, non-magnetic field route to controlling the orientation of magnetization in the material[5]–[9]. In contrast to semiconductor technology, where the deformation is statically induced via lattice mismatch, magnetic logic requires to dynamically switch between magnetization orientations and hence, requires to cycle the mechanical strain. For these applications, strong piezoelectric materials[5], [6] (PZT/PMN-PT) are required to efficiently provide the large deformations necessary to switch the magnetization of ferromagnetic materials. The combination of piezoelectric/ferromagnetic hybrids are popularly called magnetoelectric (ME) transducers[4], [9], [10]. The voltage control of magnetization delivered by ME technology is crucial to perform low power, beyond CMOS logic operations using magnetic materials.

ME transducers are also essential to the realization of spin wave computing[11]–[14], which is a different branch of magnetic logic technologies relying on encoding information in the phase of propagating spin waves. For these application, the role of the ME transducer is to perturb the magnetization order and generate spin waves by delivering lattice vibrations at frequencies above the ferromagnetic resonance (typically 1-30 GHz) [15],[16],[17]. Typical ME transducers proposed in literature for spin-wave computing applications rely on capacitive tranducers fabricated on strong piezoelectrics like PZT/PMN-PT [15],[18]. However, non-resonant operation of capacitive transducers make them fundamentally ill-equipped to be interfaced with microwave circuits. In contrast, acoustic resonators provide an established route to impedance matching and efficient transduction of energy between the mechanical and microwave domain. However, the weak piezoelectric coefficient of AlN[19] – the material conventionally used in acoustic resonators – imposes fundamental limits on the absolute magnitude of stress delivered and hence, limits the scope of its application as a ME transducer. On the other hand, the high frequency acoustic response of strong piezoelectric materials like PZT is relatively less explored owing to the assumption of large dielectric/ferroelectric losses at GHz frequencies. Thus, the characterization of the high frequency electromechanical response of strong piezoelectric materials like PZT is likely to benefit the development of ME tranducer technology.

In this work we demonstrate that high-overtone bulk acoustic resonantors (HBARs) fabricated on $Ba_{0.1}Pb_{0.9}Zr_{0.52}Ti_{0.48}O_3$ (BPZT) films can generate acoustic modes up to 15 GHz, potentially providing an unique material platform to deliver both large stress at static conditions, and high frequency perturbations. The ferroelectricity of BPZT additionally provides an electric field dependent electromechanical transduction efficiency ($K^2$) and switchability of the acoustic response. The demonstrated BPZT HBARs could also find other applications where large amplitude stress transduction at GHz frequencies is desirable like for the mechanical manipulation of spin qbits and interfacing with superconducting qbits[20]. Finally, the test structure and techniques employed in this work can be used as a generic tool to characterize the electrical and electromechanical performance of thin film piezoelectrics at high frequencies.

The paper is organized as follows. Section 2 introduces both the experimental design of the device and its suitability to the dielectric and electromechanical characterization of BPZT at high frequencies. Section 3 discusses the identification of HBAR modes from the measured electrical impedance of the device and comparision with expectations from the Mason model. Finally section 4 comments on the potential usage of our transducers for magneto-electric generation of spin-waves.

## II. EXPERIMENTAL DESIGN AND MODELS FOR EXTRACTING THE DIELECTRIC/ELECTROMECHANICAL PERFORMANCE

### *A. Material system: Barium-doped PZT thin films*

Large stress generation requires a material with strong piezoelectric coefficient, free of current leakage and with negligible losses at microwave frequencies. There is a common belief that such materials do not exist: low loss piezoelectric materials like AlN have a modest piezoelectric coefficient, while strong ferroelectric materials like PZT are expected to be lossy at microwave frequencies. In contrast to this common thinking, we will show that when properly implemented, PZT can meet our needs. We deposited Barium-doped PZT films by pulsed laser deposition in a Solmates SIP-700 system using a 10 nm thick $LaNiO_3$ buffer layer and a Pt bottom electrode. Barium incorporation in PZT reduces the grain size[21], increasing the dielectric integrity, without degrading the electromechanical performance. The Scherrer grain size, calculated from the XRD patterns, was found to be independent of the thickness and around 20-25 nm. Electron micrographs indicated that the grains were organized in a columnar fashion. The rms surface roughness of the BPZT film was as low as 6 nm. Further details about the material growth and device fabrication can be found elsewhere [22].

### *B. Geometry: concentric capacitor structures*

In order to characterize the intrinsic electromechanical properties of BPZT films at microwave frequencies, we have patterned electrodes on top of the films using a one-step photolithography followed by a lift-off of a thick top electrode. Overall, the complete stack of the device is as follows Au (930 nm)/Ti(10nm)/BPZT (400 nm)/ $LaNiO_3$ (10 nm)/Pt (70 nm)/Ti (10 nm)/ $SiO_2$ (400 nm)/Si (750 μm, substrate).

A simple concentric capacitor (CC) geometry (Fig. 1a) was chosen to preserve the rotational invariance of the BPZT material properties. The central signal pad radius, *a*, of the CC was varied from 10 to 50 μm, while, the outer ground pad radius was fixed at 80 μm. Rather small feature sizes were chosen for two mains reasons that will appear clear later: first, it allows to model the system as a lumped electrical circuit, and second because it allows a better impedance matching thus improving the sensitivity of the measurements.

### *C. Dielectric permittivity and loss tangent*

To check that the design and fabrication were satisfactory, we measured the dielectric permittivity and the loss tangent of the BPZT films using the concentric capacitors. For this we measured the scattering parameters of the capacitor in a calibrated 1-port measurement using a vector network analyser. The scattering parameters are then converted to impedance parameters using the characteristic impedance (50 Ω) of the measurement setup. The measured impedance $z = r + jx$ is separated into real ($r$) and imaginary ($x$) part. The permittivity and loss tangent extracted by comparing the impedance parameters measured on two CC with different radius ($a_1$ and $a_2$). The explicit relation is given as [23]:

$$\varepsilon_r = -\frac{\left(\frac{1}{a_1^2}-\frac{1}{a_2^2}\right) * t * (x_1 - x_2)}{(\omega\pi\varepsilon_0)\left[\left[\left(r_1 - r_2 - \frac{R_S}{2\pi}\ln\left(\frac{a_2}{a_1}\right)\right)\right]^2 + (x_1 - x_2)^2\right]},$$

$$\tan\delta = \frac{\left(r_1 - r_2 - \frac{R_S}{2\pi}\ln\left(\frac{a_2}{a_1}\right)\right)}{(x_1 - x_2)},$$

where $R_s$ is the series resistance – extracted to be 2.1 Ohms, $t$ is the thickness of BPZT (400 nm), $a_1$ and $a_2$ are the radii of two different CC, while $r_1$, $r_2$, and $x_1$, $x_2$, are respectively the real and imaginary part of the measured impedances.

The extracted permittivity and loss tangent of the 400 nm thick BPZT embodied in a CC is plotted in Fig 1(b). The dielectric permittivity was around 480 and decreased by less than 10% in the frequency range from 1-6 GHz. The dielectric loss increases with frequency as expected and was about 0.1 at 5 GHz. In addition to the monotonous capacitive response, we observed clear anomalies in the dielectric response and the loss tangent at specific frequencies (Fig. 2 and Fig. 3). These anomalies correspond to acoustic resonances in the capacitor structures.

### *D. Expected electromechanical performance of BPZT at high frequencies*

Let us model how the acoustic resonances impact the impedance spectrum. The central part of the device under the signal pad can be understood as a High overtone bulk acoustic resonator (HBAR). HBARs are composite resonators in which the acoustical modes encompass both the BPZT film and the thick substrate. The BPZT film is a small fraction of the resonator volume, and hence the quality factor Q of the composite resonator is expected to correspond essentially to the low-loss, high quality silicon substrate. Upon RF excitation, the piezoelectric BPZT pumps mechanical energy in the form of travelling acoustic waves into the substrate. The silicon substrate acts as a Fabry-Perot cavity with two mirror surfaces defined by the large acoustic mismatch at the Si/$SiO_2$ interface and the polished bottom surface respectively. Standing wave modes are created when the thickness of the substrate is an integer multiple of the acoustic wavelength. Thus, the acoustic resonant frequency of the silicon substrate determines the frequency intervals between each resonant mode or the free spectral range (FSR) of the resonator. The large thickness of silicon ($t_{Si}$) substrate implies a dense spectrum of modes spaced at the substrate resonance determined by its sound velocity ($v_{Si}$) as $v_{Si}/(2 * t_{Si})$. The FSR is expected to be ~5.5 MHz.

The amplitude of acoustic energy delivered by the piezoelectric is maximal around the thickness extensional modes of the piezoelectric film. For negligible electrode thickness, the piezoelectric resonance frequencies would be given as $v_p/(2 * t_p)$, where $v_p$ and $t_p$ are the acoustic velocity and thickness of the piezoelectric layer, respectively. However, the simple analytical formulation is not applicable for our device stack, with a thick top electrode and the isolating $SiO_2$ layer underneath. Instead, the Mason model [24] is used for a more reasonable estimate of the impedance spectrum. The Mason model [24],[25],[26] is essentially a 1-D transmission line model that accounts for the electrical impedance presented by a mechanical wave propagating in a composite piezoelectric structure like ours. The acoustic impedance and velocity of each layer in the stack (Table I) is input to derive the acoustic resonance frequencies of the entire stack. The transformation from the mechanical to the electrical domain is made by incorporating an equivalent transformer, physically representing the electromechanical performance of the piezoelectric material. Mathematically, the transformer creates a coupling of magnitude $\sqrt{\frac{d_{33}*A}{t*Y}}$ between the mechanical and electrical domains, where $d_{33}$ is the longitudinal piezoelectric coefficient, A is the area of the capacitor, $t$ is the thickness of the piezoelectric and Y is the Young's modulus of BPZT.

The material parameters (Table I) can be measured from the recursive fitting of the experimental impedance spectra with the Mason model calculation. The fitting is carried out by (i) using the permittivity and loss tangent values extracted from the CC structures to model the frequency response of the capacitor and (ii) optimizing the turns ratio of the electromechanical transformer to reproduce the acoustic response. Thus, by independently measuring the permittivity, and capacitance in CC structures, we can numerically optimize the piezoelectric coefficient and acoustic velocity of BPZT to get good agreement with the measured data.

### *E. Numerical estimation of stress generation in BPZT HBARs*

Let us now quantify the mechanical stress generated at resonance in the device. The computation of the stress is not accessible from purely electrical measurements, so we use a finite element simulation (COMSOL) to perform a numerical simulation of the complete stack and to quantify the mechanical stress generated at acoustic resonances. A frequency sweep is setup and acoustic resonances are identified as dielectric anomalies in the impedance sweep. A terminal voltage of 1V and reference impedance of 50 Ohms are used in the computation. Overtones of the fundamental thickness mode are observed as a function of the frequency. For a given geometry of the device, the frequency and the $K^2$ of the mode would determine the impedance at resonance. Maximum transduction of acoustic energy happens when the device impedance is close to 50 Ohms. Thus, the mechanical stress generated by an acoustic mode, for a constant voltage drive, would depend on the impedance of the device. As qualitatively expected, standing acoustical waves are observed in the silicon substrate at the HBAR mode frequencies. The frequencies obtained in the finite element simulation match closely those obtained from both the Mason model, and the next section will confirm that they are in line with the experimental measurements. The mechanical stress generated for the first and second thickness extensional modes of the piezoelectric film are shown in Fig 1 (c) and (d), respectively. The mechanical stress along a line cut across the HBAR structure reveals a coherent standing wave pattern across the thickness of the substrate. The oscillating stress pattern generated by the second thickness extensional mode of the BPZT (Fig 1d) has a larger amplitude and frequency than the first mode (Fig 1c) due to the better impedance matching of Mode 2, as we will confirm in the experimental impedance measurement. In order to benchmark the PZT microwave stress generator with state-of-the-art systems, we have also simulated the complete HBAR stack with AlN as piezoelectric material. The stress generated in the BPZT HBAR was found to be about an order of magnitude larger, when compared with the AlN HBAR, and can be related to the larger dielectric permittivity and piezoelectric coefficient of BPZT, as compared to AlN.

## III. RESULTS AND DISCUSSION

### *A. Spectrum of acoustic modes and corresponding electromechanical coupling*

The modulus of the complex impedance measured between 1-5.5 GHz on a CC with 50 µm radii defined on 400 nm thick BPZT film for 0 kV/cm and 250 kV/cm is shown in Fig 2. The response at 0 kV/cm is devoid of any acoustic resonance and shows the expected capacitive (*1/jωC*) response. The large permittivity of BPZT can be observed to result in a small electrical impedance for the 50-µm radius of CC at GHz frequencies. As discussed, maximum stress generation would require the impedance of the resonator to be as close as possible to the 50 Ω source impedance. Thus, further scaling of the CC geometry is required to ensure 50 Ω impedance matching of BPZT at GHz frequencies.

At an applied dc field of 250 kV/cm, four clear envelopes of HBAR modes corresponding to longitudinal extensional modes of the piezoelectric film, *m* from 1 to 4, can be identified. In the inset of Fig 2, a zoom in to mode 3 reveals that the peaks are clearly spaced around ~5.5 MHz, as expected from the acoustic velocity and thickness of the substrate. A Lorentzian fit of the peaks is made to extract the Quality factor, yielding a value of ~950, which is similar to reported values in literature for HBARs on silicon substrates [27]. This confirms our expectation that the mechanical losses in ferroelectric BPZT does not have a detrimental effect on the quality factor of the HBAR.

The Mason model estimate for the experimental measured structure is included in Fig 2, showing good correspondence with the experimental data. The piezoelectric coefficient $d_{33}$ was found to be 60-70 pm/V. This is the first estimate of the piezoelectric coefficient of BPZT at frequencies above 2 GHz, and the values are consistent with the range of values reported on thin film PZT [28],[29],[30] at low frequencies (<100 MHz). The acoustic velocity is estimated to be ~ 4800 m/s and a value of $K^2$ extracted ~12% for the first mode is much larger than AlN [19] and is close to previously measured values on PZT film bulk acoustic resonators (FBARs) [31],[32].

### *B. HBAR response versus capacitor size: electrical mismatch and physical clamping*

The impedance spectra for different radii of CC were compared (Fig. 3 (a)) to elucidate the perturbation induced in the mechanical modes by the process of physically landing the probes on the measurement pads. A clear decrease in the Quality factor of HBAR modes is observed for decreasing radius of CCs. For these structures, the clamping from the probes mechanically dampens the substrate resonances, resulting in a lossy HBAR structure corresponding to the envelope of piezoelectric thickness modes. For the 50 μm radii (Fig 3(b)), the Quality factor extracted from a Lorentzian fit of experimental data is used for Mason model estimates. However, this method is not applicable to the smaller radii due to the significant mechanical clamping of the substrate modes caused by the landed microwave probes, and the resulting damping. Nevertheless, a good correspondence with the Mason model is achieved, if we assume a lower substrate Quality factor e.g. ~200 for the 15 μm radii [Fig 3(c)].

Note that the clamping from the microwave probe reduces the resonances quality factors but not the electromechanical characteristics of the BPZT film. Thus, even the capacitors with the smallest radii (15 μm) could be used to probe the characteristics of the film. Using a smaller capacitor area allows better impedance matching to 50 Ohms, a higher frequency RC cut-off, and lower leakage current and. Note that 50 Ohm impedance matched BPZT capacitors would require sub-10 μm capacitor surfaces, the artefact induced by the mechanical clamping by the microwave probes can be problematic. To benefit from the best of the ferroelectric material, it is desirable to laterally separate the HBAR mode from its electrical contact. A possible route is the use of an overlap capacitor geometry [33].

### *C. Switchable HBAR response though ferroelectricity*

In addition to its large piezoelectric coefficients, a further advantage of BPZT is that, being a ferroelectric material, exhibits a field-tenable electromechanical coupling the effective electromechanical coupling is field tuneable thanks to the intrinsic ferroelectricity. The field dependence of the piezoelectric modes is shown in Fig. 4a with a frequency-electric field colour coded map with the value of the real part of the device impedance. Clear regions correspond to the case when the ferroelectric is in unpoled, non-piezoelectric state. For each resonant mode, the amplitude of the real part of the impedance increases with DC bias, indicating an increase in the transduction of acoustic energy. A gradual broadening in frequency space is also observed for increasing electric field, indicating a corresponding increase in the $K^2 = \frac{d_{33}^2 * Y}{\varepsilon_r * \varepsilon_0}$.

Let us understand the origin of this increased response. The capacitance and piezoelectric coefficient extracted from the Mason model are summarized in Fig. 4b for the same structure. As expected for a ferroelectric film, a large decrease in dielectric permittivity is observed as a function of the applied electric field. The piezoelectric coefficient saturates at larger fields and shows variation only around coercive fields due to the switching of the ferroelectric polarization. Thus, the continuous increase in the $K^2$ observed in Fig. 4a can be explained as originating from the sole decrease in the dielectric permittivity for a fixed piezoelectric coefficient. This bias induced control of the $K^2$ provides an elegant route for controlling the electromechanical performance of BPZT HBARs stress generators. In contrast, no such route exists to control the performance of current state-of-the-art HBARs realized with simple non-ferroelectric piezoelectric materials like AlN.

The impedance spectra for different radii of CC were compared (Fig. 3 (a)) to elucidate the perturbation induced in the mechanical modes by the process of physically landing the probes on the measurement pads. A clear decrease in the Quality factor of HBAR modes is observed for decreasing radius of CCs. For these structures, the clamping from the probes mechanically dampens the substrate resonances, resulting in a lossy HBAR structure corresponding to the envelope of piezoelectric thickness modes. For the 50 μm radii (Fig 3(b)), the Quality factor extracted from a Lorentzian fit of experimental data is used for Mason model estimates. However, this method is not applicable to the smaller radii due to the significant mechanical clamping of the substrate modes caused by the landed microwave probes, and the resulting damping. Nevertheless, a good correspondence with the Mason model is achieved, if we assume a lower substrate Quality factor e.g. ~200 for the 15 µm radii [Fig 3(c)].

### *D. High frequency capability of BPZT HBARs*

The excitation of acoustic modes at microwave frequencies in BPZT films is commonly not expected. Indeed, it is generally anticipated that at these frequencies, the large tanδ of BPZT resulting from the large permittivity would fundamentally prohibit acoustic actuation. In our experiments, we found that while the extracted Tanδ >0.1 at 5 GHz, is very large when compared to AlN << 0.01 or $LiNbO_3$ < 0.05, it does not impede the excitation of HBAR modes. This arises from the fact that the quality factor of the resonator is dominated by the substrate response instead of the sole PZT response. To illustrate this point, we plot in Fig. 5 the high frequency impedance spectra measured on 15 µm CC. Clear signatures of acoustic modes can be observed up to 15 GHz. Furthermore, all observed acoustic modes were found to be switchable and controllable with the magnitude of the applied DC bias. Thus, the observation of acoustic modes in BPZT up to 15 GHz, shows promise for using strong electromechanical materials for high frequency stress transduction.

## IV. Prospects for spin-wave manipulation

The motivation of our work is to demonstrate that strong piezoelectric materials like BPZT which are already used as ME transducers for dc functionalities could also simultaneously excite magnetization dynamics at microwave frequencies through inverse magnetostriction in a magnetic film by coupling to the HBAR modes. Magnetoelastic coupling of thickness extensional HBAR modes to in-plane magnetized films has already been demonstrated by canting the piezoelectric polarization of ZnO HBARs [34]. For the sake of illustration, let us assume a similar canting of the ferroelectric polarization can be realized in BPZT HBARs and compare their prospects as SW transducers. In the case of BPZT HBARS, the mechanical strain upon 1V excitation is estimated to reach values up to $\varepsilon_{zz}=10^{-3}$. Assuming a magneto-elastic coupling coefficient of $B_1$=25.4 T for Ni [35], the strain would translate to a dynamic effective magnetic field of 25 mT that can be used for a properly chosen misalignment between the magnetization and the strain axis. This applies if the magnetic material is positioned directly under the central pad of the CC and experiences a uniform $\varepsilon_{zz}$ strain field. Alternatively, the magnetic material can be coupled evanescently, by positioning it in lateral proximity to the CC, to harness magnetoelastic coupling to other components of the stress tensors. In any case, the strength of the RF magneto-elastic field can be about two orders of magnitude larger than the Oersted fields generated by inductive microwave antennas based SW transducers for a 1 V drive [36]. Furthermore, the large frequency span of operation enabled displayed by the BPZT HBARs implies the generation of mechanical strain with a large bandwidth of wavevectors ranging from 1 radians/um to 20 radians/um – vastly exceeding the capabilities of conventional microwave antennas [36]. Finally, HBARs are technologically simple to fabricate, and the longitudinal mechanical stress generated could also be conveniently coupled to SW devices fabricated on the other side of the substrate, as demonstrated in [37].

## V. Conclusion

In summary, we have reported the first experimental demonstration of switchable, bias controlled HBAR modes excitable up to 15 GHz in BPZT films deposited on silicon. The measured response was fully comparable with a simple 1D Mason model estimate. The robust high frequency electromechanical performance of BPZT suggests a route to using the same material platform for large amplitude stress generation at both DC and GHz frequencies, as desired for magnetoelectric transducers technologies.

## Acknowledgment

This work was funded through the FETOPEN-01-2016-2017 - FET-Open research and innovation actions (CHIRON project: Grant agreement ID: 801055).

**Table 1 Material properties used for the Mason model calculations**

| **Layer** | **Thickness (nm)** | **Velocity (m/s)** | **Acoustic Impedance (M.Ohms)** |
|---|---|---|---|
| Silicon substrate | 750*1000 | 9400 | 19.68 |
| Silicon oxide | 400 | 5800 | 12.85 |
| Combined Bottom electrode (Pt/LNO) | 92 | 3342 | 69.41 |
| BPZT | 400 | 4800 | 30.1 |
| Top electrode (Ti/Au) | 930 | 3356 | 63.8 |

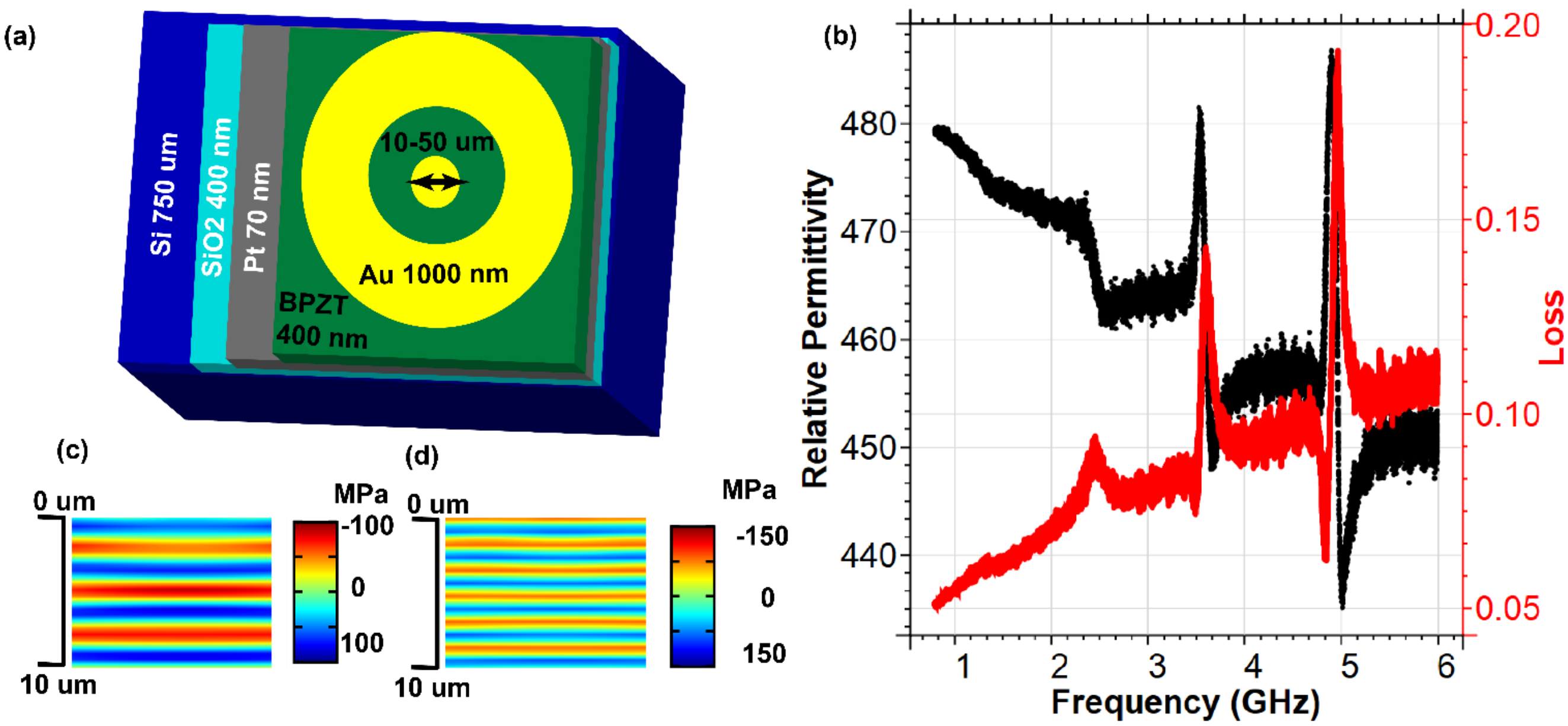

**Figure 1** (a) Simplified cross-section of the BPZT HBAR structure and illustration of concentric capacitor design; (b) the dielectric permittivity and loss extracted from the concentric capacitor structures; finite element model simulation of stress generated in the first (c) and second (d) longitudinal extensional modes for a 400 nm thick BPZT HBAR. The vertical scale starts at the SiO2/Si interface below the central pad of the concentric capacitor.

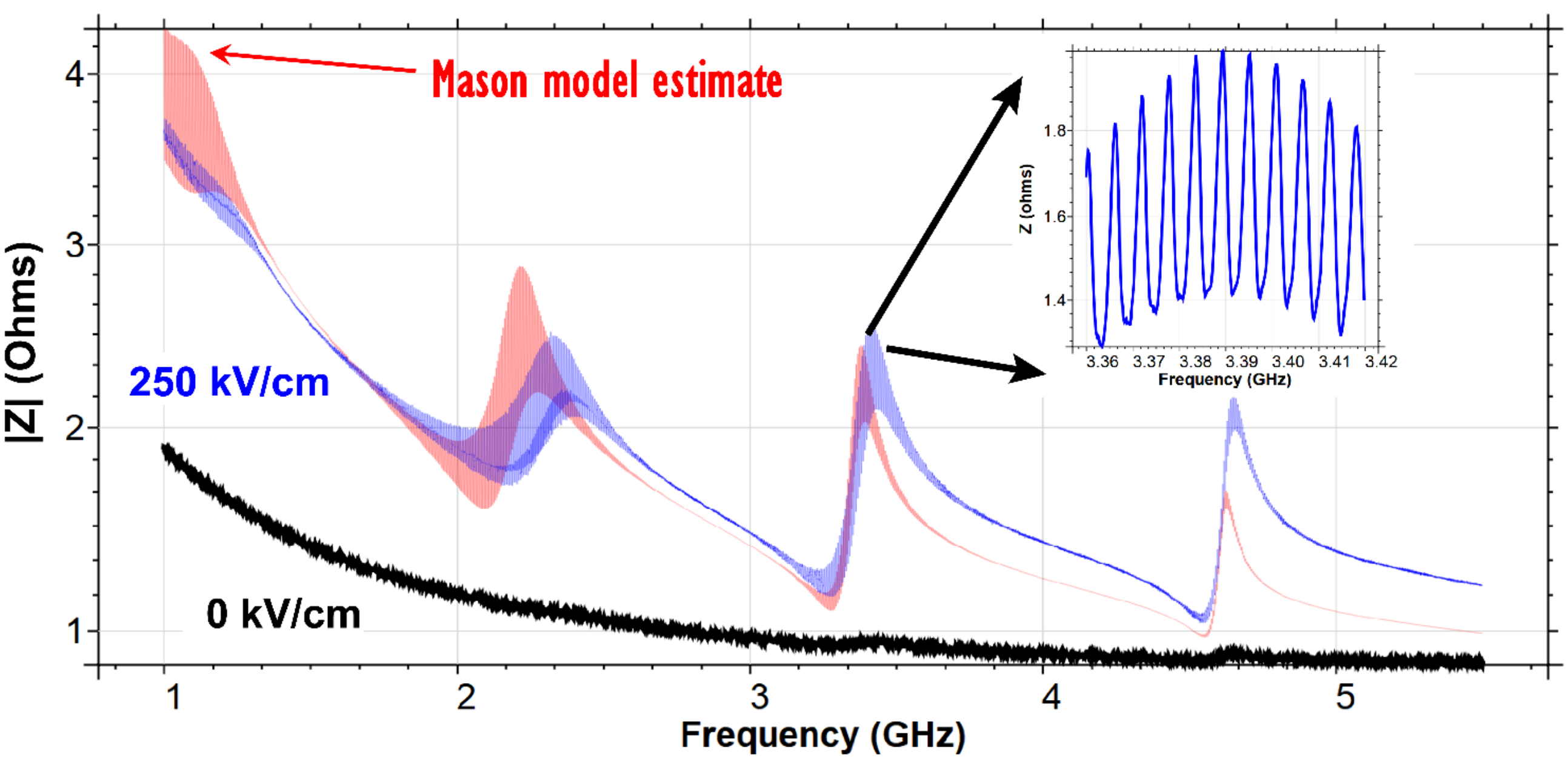

**Figure 2:** Spectrum of the modulus of the impedance for a 400 nm thick BPZT HBAR based on a 50 µm concentric capacitor; four mode envelopes corresponding to increasing overtones of the fundamental thickness mode of the PZT film are observed. The inset zooms into one envelope to reveal peaks spaced at the resonant frequency of the substrate~5.5 MHz. Peaks share the same quality factor of 950.

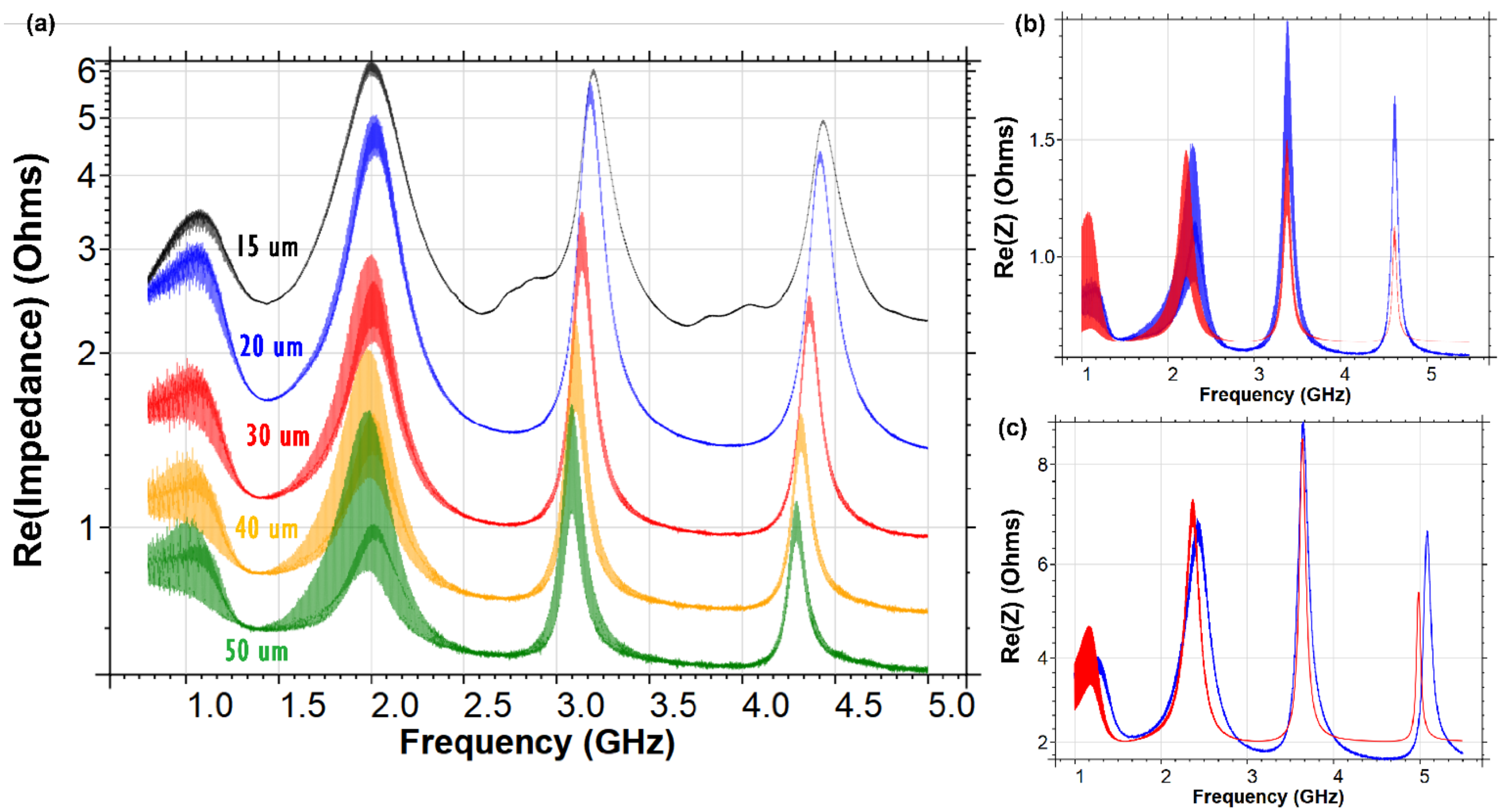

**Figure 3**(a) The real part of the impedance for a 400 nm thick BPZT film is plotted for different radii of concentric capacitors for 50 to 15 µm; (b) The experimental results (blue curve) and Mason model estimate (red curve) show good agreement for (b) 50 µm using a quality factor of 950 and (c) for 15 µm CC using a quality factor of 200.

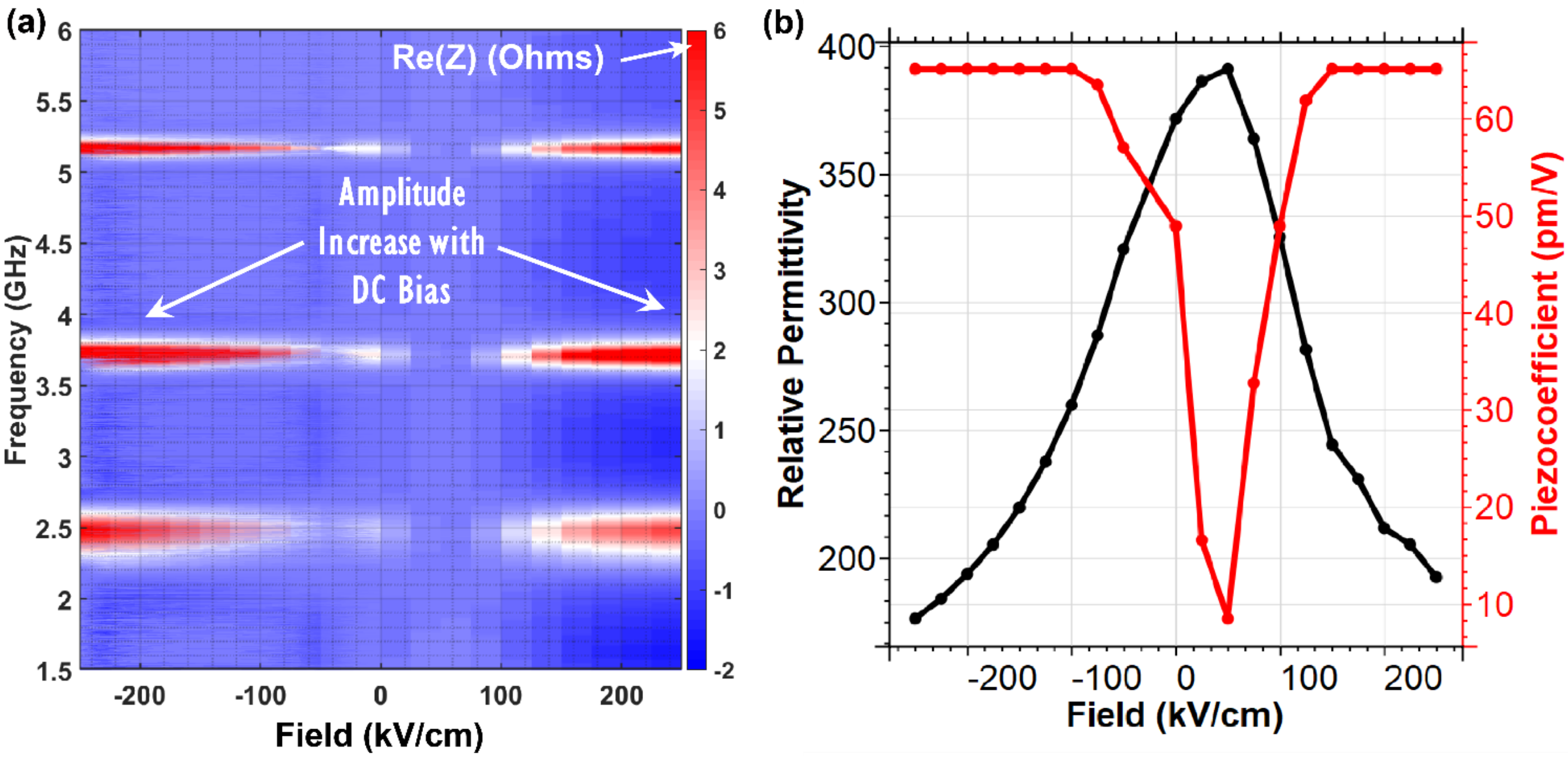

**Figure 4**: (a) The DC bias dependence of the resonances are shown by plotting the frequency-electric field map with the color representing the magnitude of the real part of the impedance for a 15 μm radius device on a 400 nm thick BPZT; (b) The permittivity extracted from the MBVD model is plotted as a function of the electric field and the extracted piezoelectric coefficient is also plotted for comparison.

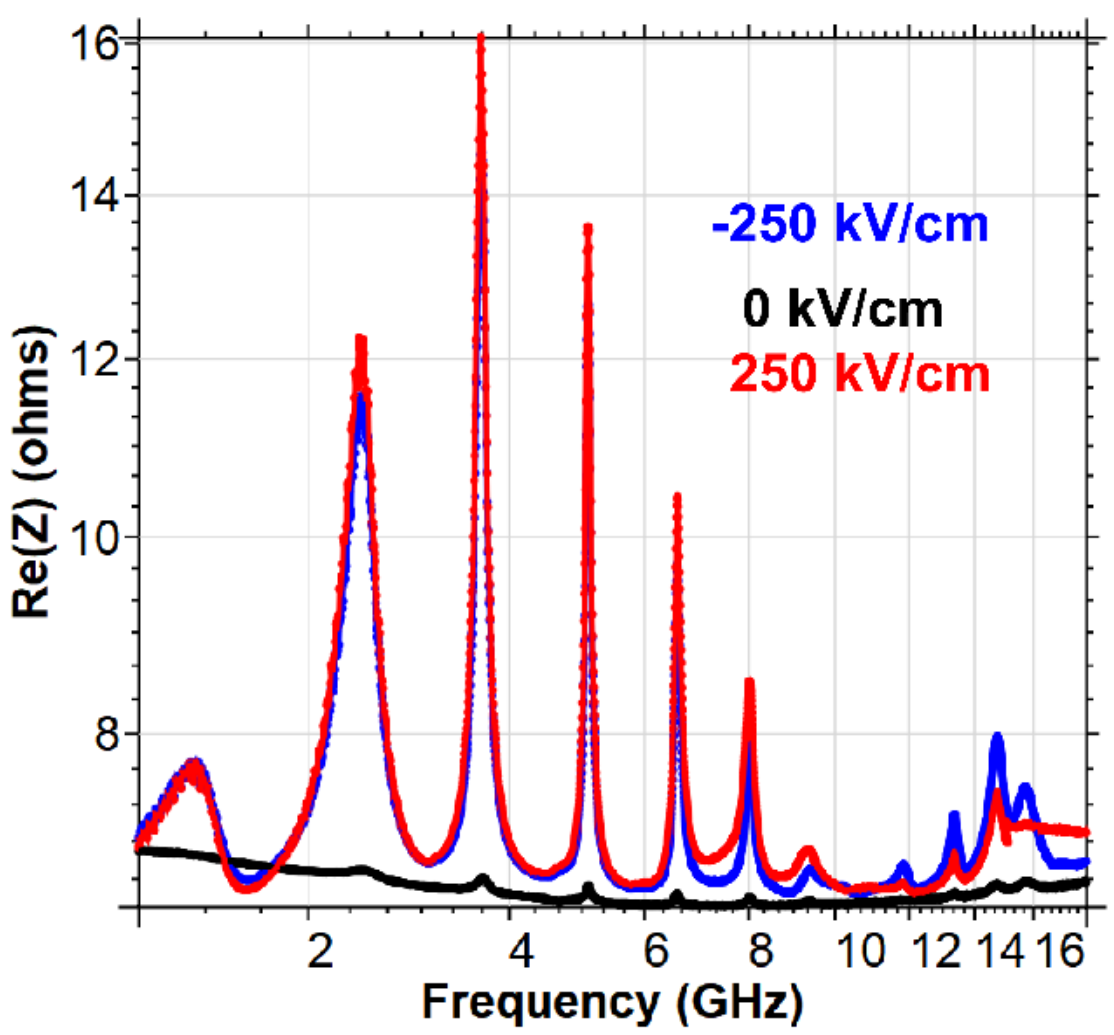

**Figure 5:** The real part of the impedance for a 10-µm radius of CC is plotted up to 16 GHz for 400 film for positive, negative and zero electric fields.